\documentclass[showpacs,twocolumn,aps,prl]{revtex4}
\usepackage{amsmath}
\usepackage{amsfonts}
\usepackage{amssymb}
\usepackage{amsthm}
\usepackage{graphicx}

\begin{document}

\title{Magnetization reversal through synchronization with a microwave}
\author{Z. Z. Sun}
\affiliation{Physics Department, The Hong Kong University of
Science and Technology, Clear Water Bay, Hong Kong SAR, China}
\author{X. R. Wang}
\affiliation{Physics Department, The Hong Kong University of
Science and Technology, Clear Water Bay, Hong Kong SAR, China}
\date{\today}

\begin{abstract}
Based on the Landau-Lifshitz-Gilbert equation, it can be shown
that a circularly-polarized microwave can reverse the magnetization
of a Stoner particle through synchronization. In comparison
with magnetization reversal induced by a static magnetic field,
it can be shown that when a proper microwave frequency is used the
minimal switching field is much smaller than that of precessional
magnetization reversal. A microwave needs only to overcome the
energy dissipation of a Stoner particle in order to reverse
magnetization unlike the conventional method with a static magnetic
field where the switching field must be of the order of magnetic
anisotropy.
\end{abstract}
\maketitle
{\it Introduction--}Magnetization reversal of single-domain
magnetic nano-particles (Stoner particles)\cite{Hillebrands} is
of significant interest in magnetic data storage and spintronics.
Finding an effective way to switch magnetization from one state
to another depends on our basic understanding of magnetization
dynamics. Magnetization can be manipulated by laser\cite{Bigot},
a polarized electric current\cite{Slon,current} or a magnetic
field\cite{He}. An important issue in magnetization reversal is
the minimal switching field. Magnetization reversal using a
static magnetic field\cite{He,Back,Schum,xrw}
or polarized electric current\cite{Slon,current} has received
close attention in recent years but there has been little
investigation on microwave induced magnetization reversal.
Thirion {\it et al.}\cite{grenoble} made probably the first
attempt in this direction. It was shown that a dramatic reduction
of the minimal switching field is possible by applying a small
radio-frequency (RF) field pulse (the decrease in the static
field is much larger than the amplitude of the RF-field).
In this paper, it is shown that a circularly-polarized microwave
on its own can induce magnetization reversal. The minimal switching
field depends on the microwave frequency. It can be shown that the
minimal switching field is at a minimum at an optimal frequency.
This optimal frequency is near the natural precession frequency at
which the particle experiences the largest dissipation.
At this optimal frequency, the switching field strength
can be much smaller than the so-called Stoner-Wohlfarth (SW)
limit\cite{Stoner} and precessional magnetization switching
field\cite{He,Back,Schum} for a static magnetic field.
Far from the optimal frequency, the switching field can be
larger than the SW-limit.

\begin{figure}[htbp]
 \begin{center}
\includegraphics[width=7.cm, height=5.cm]{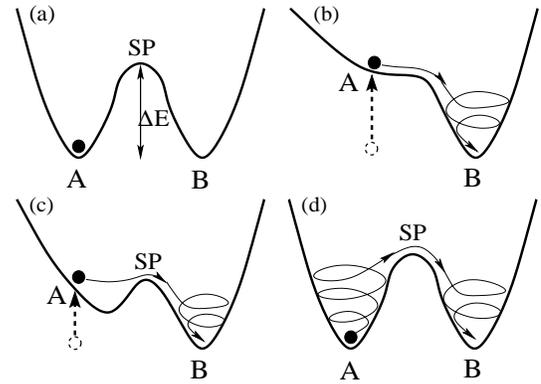}
 \end{center}
\caption{\label{fig1} Energy surface of a uniaxial magnetic
particle in various schemes. $SP$ denotes the saddle point between
two minima. (a) In the absence of magnetic fields: $A$ and $B$ are
the two minima, separated by a potential barrier $\Delta E$. (b)
At the SW-limit: Target state $B$ is the only minimum. (c)
Precessional magnetization reversal: The particle energy at $A$ is
higher than that at $SP$ so that it can pass through $SP$ under
its own dynamics. (d) New strategy: The system synchronizes its
motion with a microwave, and climbs over the potential barrier to
reverse its magnetization.}
\end{figure}

The minimal switching field was first studied by Stoner and
Wohlfarth\cite{Stoner}. The SW-limit is the field at which the
energy minimum around the initial state is destroyed and the
target state is the only minimum\cite{He,Schum,xrw,Back},
as illustrated in Fig. \ref{fig1}(a)-(b). In the absence
of magnetic fields, two energy minima [$A$ and $B$ in Fig.
\ref{fig1}(a)], separated by a potential barrier $\Delta E$,
are along the easy-axis of a magnetic particle. At the SW-limit,
the original minimum near the initial state $A$ disappears
[Fig. \ref{fig1}(b)], and the particle will end up at its unique
minimum near the target state $B$. Recent theoretical and
experimental studies\cite{He,Back,Schum} have shown that
the minimal switching field could be smaller than the SW-limit.
The reason has been explained earlier\cite{xrw}. As illustrated in Fig. \ref{fig1}(c),
magnetization reversal can occur even when the minimum around $A$
exists. The reversal can happen as long as the particle energy at
$A$ is higher than that at the saddle point $SP$, and the particle
can pass through $SP$ under its own dynamics. It can be shown that
the minimal switching field is of the order of the potential barrier
$\Delta E$\cite{xrw}.

{\it Fundamental differences between time-independent and
time-dependent fields--}The microwave induced magnetization
reversal is fundamentally different from that of a static magnetic
field because a static field is not an energy source whilst a
microwave can. This can be seen from the dynamic equation
governing the evolution of a single-domain magnetic nano-particle.
For a particle with a magnetization of $\vec{M}=\vec{m}M_s$,
$\vec{m}$ satisfies the Landau-Lifshitz-Gilbert (LLG)
equation\cite{xrw,Landau},
\begin{equation}
(1+\alpha^2)\frac{d\vec{m}}{dt} = - \vec{m} \times \vec{h}_{t}-
\alpha \vec{m} \times (\vec{m} \times \vec{h}_{t}), \label{LLG}
\end{equation}
where $M_s$ is the saturated magnetization of the particle, and
$\alpha$ is a dimensionless damping constant.
The total field, measured in unit of $M_s$,  comes from an
applied magnetic field $\vec{h}$ and the internal effective
field $\vec{h}_i$ due to the magnetic anisotropy $w(\vec{m},
\vec{h})$, $\vec{h}_{t}=-\nabla_{\vec{m}}w(\vec{m} ,\vec{h})
=\vec{h}_i+\vec{h}$.
In Eq.~\eqref{LLG}, time $t$ is in unit of $(|\gamma|M_s)^{-1}$
with $|\gamma|=2.21 \times 10^5(rad/s)/(A/m)$ being the
gyromagnetic ratio. From Eq.~\eqref{LLG}, the energy change rate
for the particle can be obtained\cite{xrw2}
\begin{equation}
\frac{dw}{dt} =\vec{m}\cdot \dot{\vec{h}}- \frac{\alpha}
{1+\alpha^2} |\vec{m}\times \vec{h}_{t}|^2.
\label{energyrate}
\end{equation}
The second term due to the damping is always negative, while
the first term due to the external magnetic field can be
either positive or negative if the field varies with time.
Thus, a time-dependent magnetic field can be both energy
source and energy sink.

{\it New strategy--}Having explained that a microwave can be
an energy source, the {\it synchronization} phenomenon of
nonlinear dynamic systems\cite{szz} can be used to reverse the
magnetization of a Stoner particle by shining the particle with
only a circularly polarized microwave. If the propagating
direction of the microwave is along the particle's easy-axis (the
magnetic field rotates around the easy-axis with the microwave
frequency), the particle's magnetization in a synchronized
motion precesses around the axis with the microwave frequency.
As illustrated in Fig. \ref{fig1}(d), the magnetization
starting from its initial minimum $A$ obtains energy from the
microwave, and eventually reaches its synchronized state.
If the synchronized state is over the saddle point $SP$
and on the side of minimum $B$, the magnetization reversal
is realized when the microwave radiation is turned off
because magnetization will end up at minimum $B$ through the
usual ringing effect\cite{He}. It is known that a nonlinear
dynamic system under an external periodic field may undergo
a non-periodic motion other than synchronization\cite{szz}.
In general, the reversal criterion is: {\it The magnetization is
reversed if the system can cross the saddle point $SP$ in Fig. 1}.


To demonstrate the feasibility of the new strategy,
an uniaxial magnetic anisotropy is considered,
\begin{equation}
w(\vec{m},\vec{h}=0)=-k m_x^2, \label{uniaxial}
\end{equation}
where $k>0$ measures the anisotropy strength.
Without losing the generality, $k$ (used as a scale for the
field strength according to Eq. \eqref{LLG}) shall be set to $1$.
The easy-axis is chosen to be along the x-axis rather than the
z-axis because the north and south poles are singular in spherical
coordinates, and it is more convenient to locate the minima $A$
and $B$ (Fig. 1) away from the singularities.

\begin{figure}[htbp]
 \begin{center}
\includegraphics[width=7.cm, height=5.cm]{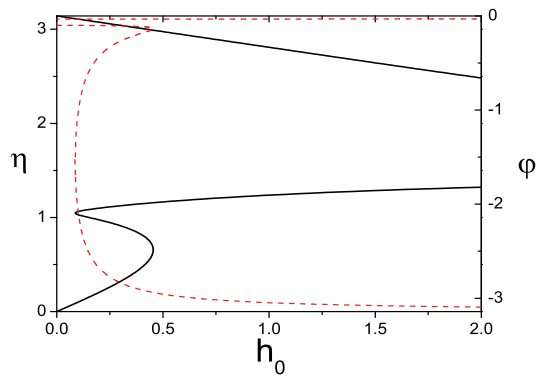}
 \end{center}
\caption{\label{fig2}(color online) Graphic demonstration of
multiple synchronization solutions. The solid lines are from
Eq.~\eqref{eta} and the dashed lines are from Eq.~\eqref{varphi}.
The graph is plotted at $\alpha=0.1$ and $\omega=1$. }
\end{figure}

{\it Multiple synchronization solutions--}Under a
circularly-polarized microwave of amplitude $h_0$ and frequency
$\omega$
\begin{equation}
\vec{h}(t)=h_0[\cos(\omega t) \hat{y}+ \sin(\omega t) \hat{z}],
\label{field}
\end{equation}
the synchronized motion is
\begin{equation}
\vec{m}(t)=\cos\eta\hat{x} + \sin\eta [\cos(\omega t+\varphi)
\hat{y}+\sin(\omega t +\varphi) \hat{z}],
\end{equation}
where $\eta$ (a constant of motion) is the precessional angle
between $\vec{m}$ and the x-axis. $\varphi$ is the {\it locking
phase} in the synchronized motion. Substitute Eqs. (3), (4), and
(5) into Eq. (1), $\eta$ and $\varphi$ satisfy
\begin{align}
&\sin\eta \sqrt{\alpha^2 \omega^2 +(2-\omega/\cos\eta)^2}=h_0,
\label{eta}\\
&\sin \varphi =-\alpha \omega \sin\eta/h_0,\label{varphi}
\end{align}
where $\eta\in [0,\pi]$. For fixed ($h_0, \omega, \alpha$),
$\eta$ and $\varphi$ may have multiple solutions. As illustrated
in Fig.~\ref{fig2}, the solutions of $\eta$ (solid lines) are
plotted as a function of $h_0$ for $\omega=1$ and $\alpha=0.1$.
The dashed lines denote the corresponding $\varphi$. Multiple
solutions of $\eta, \varphi$ are evident. For example, there are
4 solutions of $\eta$ when $h_0 \in [0.09, 0.45]$.
Numerically, it can be shown that two solutions around $\eta=1$
(in between) are unstable, while the other two near $\eta=0,\pi$
are stable. Thus, the system shall eventually end up at one of
the two stable solutions. Which one the system will choose
depends on the initial condition. For a given initial condition
[$\vec{m}(0)=\hat x$ in this study], the system picks the solution
near $\eta=\pi$ when $h_0$ is larger than a critical value called
the minimal switching field. According to our reversal criterion,
the magnetization is reversed through synchronization.


\begin{figure}[htbp]
 \begin{center}
\includegraphics[width=7.cm, height=5.cm]{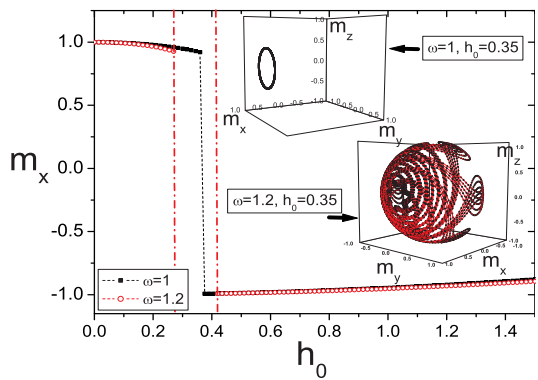}
 \end{center}
\caption{\label{fig3}(color online) $m_x$ of synchronized motion
vs. $h_0$ for $\alpha=0.1$ and $\omega=1$ (filled squares); $1.2$
(open circles). Non synchronized motion when $h_0\in [0.27, 0.42]$
(between two dash-dotted lines) is found for $\omega=1.2$. Upper
inset: Long time trajectory of $\vec m(t)$ for $\omega=1$ and
$h_0=0.35$. Lower inset:  Long time trajectory of $\vec m(t)$ for
$\omega=1.2$ and $h_0=0.35$.}
\end{figure}

{\it Numerical verification of synchronized and non-synchronized
motion--}A nonlinear dynamic system under an external periodic
field may undergo motion other than synchronized.
Unfortunately, a non-synchronized motion is, in general, hard to
define analytically. Usually, reliance must be placed on the
numerical method. In terms of the LLG equation under a
circularly-polarized microwave of Eq. \eqref{field}, it is
straight forward\cite{xrw} to calculate numerically $\vec m(t)$
starting from $\vec m(0)=\hat x$. The upper inset of Fig.
\ref{fig3} is the trajectory of $\vec m(t)$ after long time in
$m_xm_ym_z$ space for $h_0=0.35; \omega =1;$ and $\alpha=0.1$. A
simple closed loop in a plane parallel to the yz-plane indicates
that this is a synchronized motion. Alternatively, the lower
right inset of Fig. \ref{fig3} is the long time trajectory of
$\vec m(t)$ for $h_0=0.35; \omega=1.2;$ and $\alpha=0.1$. Its
motion is very complicated, corresponding to a non-synchronized
motion. It is found that whether the motion is synchronized or
not it is sensitive to the microwave frequency. For example, all
motions for $\omega=1$ are synchronized while both synchronized
and non-synchronized motions are possible for $\omega=1.2$.
The motion is non-synchronized for $h_0$ in the range of $[
0.27,0.42]$ while it is synchronized for other values of $h_0$.
Fig. \ref{fig3} is $m_x$ of synchronized motions as a function
of $h_0$ for $\omega=1$ and $1.2$.

\begin{figure}[htbp]
 \begin{center}
\includegraphics[width=7.cm, height=5.cm]{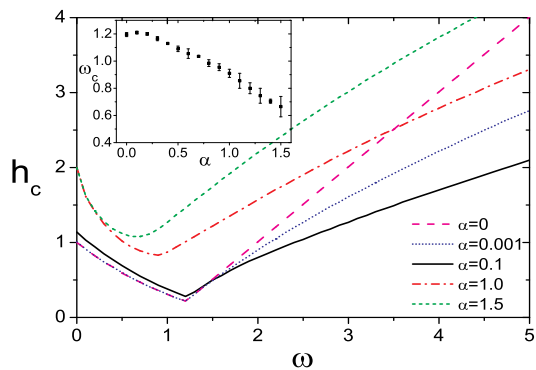}
 \end{center}
\caption{\label{fig4}(color online) The minimal switching field
$h_c$ vs. $\omega$ for various damping constant $\alpha=0$;
$0.001$; $0.1$; $1$; and  $1.5$. Inset: The optimal frequency
$\omega_c$ vs. $\alpha$. }
\end{figure}

{\it Optimal microwave frequency--}Using the reversal criterion
given earlier, it can be shown from Fig. \ref{fig3}, that the
minimal switching field $h_c$ is about $0.375$ for $\omega=1$
because $m_x$ in the synchronized motion is negative when $h_0>
h_c$. For $\omega=1.2$, the minimal switching field takes a value
at which the magnetization undergoes a non-synchronized motion.
Numerically, it can be shown that $\vec m$ crosses the yz-plane
when $h_0\ge 0.285$. Thus, the minimal switching field is
determined as $h_c=0.285$ for $\omega=1.2$. The reason that the
value of the minimal switching field is so sensitive to the
microwave frequency is because a switching field, as illustrated
in Fig. 1(d), needs to overcome the dissipation which is related
to the motion of the magnetization (see the LLG equation).
To reveal the frequency dependence of the minimal switching
field, Fig. \ref{fig4} shows the minimal
switching field $h_c$ vs. the microwave frequency $\omega$ for
various $\alpha=0$; $0.001$; $0.1$; $1$; and $1.5$. $\omega=0$
corresponds to the case of a static field along the y-axis.
The curve of $\alpha=0$ intersects the $h_c$-axis at $h_c=1$ which
agrees with the exact minimal switching field $h_c=1$\cite{xrw}.
The intersections of all other curves of $\alpha\ne 0$,
are the same as those with a static field\cite{He,xrw}.
When $\alpha\geq 1$, it becomes the SW-limit $h_c=2$.
For a given $\alpha$, Fig. \ref{fig4} shows the existence of an
optimal microwave frequency, $\omega_c$, at which the minimal
switching field is the smallest. Far from the optimal frequency,
the minimal switching field can be larger than the SW-limit.
The inset of Fig. \ref{fig4} is $\omega_c$ vs. $\alpha$.
The optimal frequency is near the natural precessional
frequency at which the dissipation is a maximum.

\begin{figure}[htbp]
 \begin{center}
\includegraphics[width=7.cm, height=5.cm]{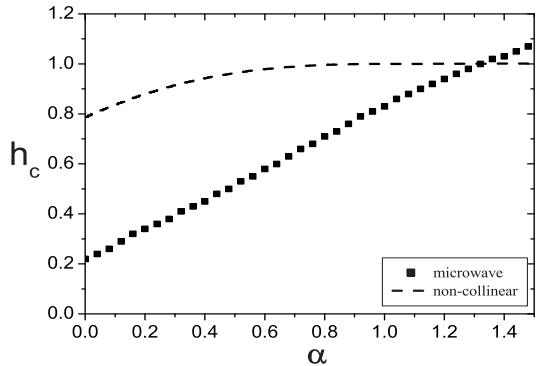}
 \end{center}
\caption{\label{fig5} $h_c$ vs $\alpha$ for the uniaxial model of
Eq. \eqref{uniaxial} under different reversal schemes. Square
symbols are the numerical results of $h_c$ at the optimal
frequency in the present strategy with a circularly-polarized
microwave. The dashed line is $h_c$ under a non-collinear static
field of $135^{\circ}$ to the easy-axis. It saturates to the
SW-limit beyond $\alpha=1$\cite{xrw}.}
\end{figure}

{\it Switching field as a function of dissipation--}From above
discussions, it can be seen that the minimal switching field is
a minimum at the optimal frequency $\omega_c$. The squares
symbols in Fig. \ref{fig5} are the minimal switching fields at
$\omega_c$ with different damping constant $\alpha$ for the
uniaxial model of Eq. \eqref{uniaxial}. They follow
approximately the line of $h_c \approx 0.23+0.58\alpha$.
This approximate linear relation is related to the fact that the
damping (field) is proportional to $\alpha$.
For comparisons, the minimal switching fields of a precessional
magnetization reversal under a static magnetic field and the
SW-limit are also plotted in Fig. \ref{fig5}.
It can be seen that, for small damping,  the smallest (at the
optimal frequency) minimal switching field can be much smaller than
that in the precessional magnetization reversal. For large
damping, the switching field can be larger than the SW-limit.

{\it{Discussion and conclusions--}}It is important to compare the
present strategy with other strategies involving time-dependent fields.
Firstly, the current scheme is fundamentally different from that in
the experiment of Thirion {\it et al.}\cite{grenoble} in several
aspects. 1) A circularly-polarized microwave of fixed frequencies
is the only switching field in the new scheme. While in reference
\cite{grenoble}, a linear polarized RF field is used as an
additional external field to reduce the main static switching
magnetic field. 2) For a Co particle of $H_i=10^5A/m $\cite{Back},
the optimal frequency is about order of $10GHz$ rather than GHz
employed in reference \cite{grenoble}. At $GHz$, Fig. \ref{fig4}
shows that the switching field would be too large to have any
advantage over a static field. The current scheme is also very
different from that in reference~\cite{xrw2} in many aspects. 1)
The time-dependent field in reference~\cite{xrw2} is used as a
ratchet that should be adjusted with the motion of magnetization.
In contrast, the present scheme is based on the synchronization
phenomenon in nonlinear dynamics such that a circularly-polarized
microwave of fixed frequencies is used and the magnetization
motion is synchronized with the microwave in the reversal process.
2) The switching field in reference~\cite{xrw2} is in general
non-monochromatic and very complicated, requiring a precise
control of time-dependent polarization. Thus, it would be a great
challenging to generate such a field. Alternatively, the current
scheme is much easier to implement and it could be technologically
important.

In conclusion, a circularly-polarized microwave induced
magnetization reversal is proposed. The proposal is based on the
facts that a microwave can constantly supply energy to a Stoner
particle, and the magnetization motion can be synchronized with
the microwave. It can be demonstrated that a Stoner particle under
the radiation of a circularly-polarized microwave can indeed move
out of its initial minimum and climb over the potential barrier.
The switching field at the optimal microwave frequency will be
much smaller than the SW-limit and that of the precessional
magnetization reversal for small damping.

{\it{Acknowledgments}--}This work is supported by UGC, Hong Kong,
through RGC CERG grants.

\end{document}